\documentclass[aoas,nameyear,dvips]{arximspdf}

\doi{10.1214/10-AOAS427}
\volume{4}
\issue{4}
\pubyear{2010}
\firstpage{1621}
\lastpage{1633}

\makeatletter
\renewcommand{\citep}[1]{[\citet{#1}]}
\makeatother

\begin{document}
\begin{frontmatter}

\title{Remembering Leo Breiman\thanksref{T1}}
\runtitle{Remembering Leo Breiman}

\begin{aug}
\author{\fnms{Adele} \snm{Cutler}\corref{}\ead[label=e1]{adele.cutler@usu.edu}}
\runauthor{A. Cutler}
\affiliation{Utah State University}
\address{Department of Mathematics\\
 \quad  and Statistics\\
Utah State University\\
UMC 3900\\
Logan, Utah 84322-3900\\
USA\\
\printead{e1}} 
\end{aug}
\thankstext{T1}{For Jessica, Rebecca and Mary Lou Breiman.}
\received{\smonth{10} \syear{2010}}

%
\begin{abstract}
Leo Breiman was a highly creative, influential researcher with a
down-to-earth personal style and an insistence on working on important
real world problems and producing useful solutions.
This paper is a short review of Breiman's extensive contributions to
the field of applied statistics.
\end{abstract}

%

\end{frontmatter}

\section{Introduction}

How many theoretical probabilists walk away from a tenured faculty
position at a top university and set out to make their living as
consultants? How many applied consultants get hired into senior faculty
positions in first-rate research universities? How many professors with
a fine reputation in their field, establish an equally fine reputation
in a \textit{different} field, \textit{after} retirement? Leo Breiman did
all of these things and more. He was an inspiring speaker and a
convincing writer, doing both with seemingly boundless enthusiasm, in
an unpretentious, forthright manner that he called his ``casual,
homespun way.''
He was intelligent and thought deeply about research. But there are a
number of bright, talented statisticians. What made Breiman different?
For one thing, he was willing to take risks. By and large,
statisticians are not great risk-takers. We tend not to stray too far
from what we know, tend not to tackle problems for which we have no
tools, tend to adopt or adapt existing ideas instead of coming up with
completely new ones. Linked to this willingness to take risks was
Breiman's unusual creativity. It was not a wild, off-the-wall
creativity---it was grounded in a sound knowledge of theoretical
principles and directed by an intuition gained by working intensively
with data, along with a generous dose of common sense. Breiman was
driven by challenging and important real-data problems that people
cared about. He didn't spend time publishing things just because he
could, filling the gaps just because they were there. Lastly, he was
tenacious. He would not give up on a problem until he, or someone else,
got to the bottom of what was going on.

Some of Breiman's ideas have advanced the field in and of themselves
(e.g., bagging, random forests) while others have contributed more
indirectly (e.g., Breiman's nonnegative garrote \citep{garrote} inspired
the lasso \citep{lasso}).
Although his joint work tree-based methods [\citet{CART}] was arguably
his most important contribution to science, he viewed random forests as
the culmination of his work. I consider myself privileged to have been
able to work with Leo Breiman for almost 20 years, as his student,
collaborator and friend, and I'm honored to have been asked to write
this review of his contributions to applied statistics. I have divided
the paper into roughly chronological sections, but these have
considerable overlap and are intended to be organizational rather than
definitive. I kept biographical details to a minimum; those interested
in a biography are referred to \citet{Olshenbio}. I do not feel
qualified to discuss Breiman's work on the 1991 Census adjustment
[\citet{census}] and have omitted a few other isolated pieces of work
such as \citet{censreg}; \citet{Varkern}; \citet{pasting}; \citet{global}.

\section{Early work}
Breiman was born in New York City in 1928 and
educated in California, receiving his Ph.D. in mathematics from UC
Berkeley in 1954. He earned tenure as a probabilist in the UCLA
Mathematics Department but soon after, he ``got tired of doing theory
and wanted something that would be more exciting'' (personal
communication) so he resigned. At this time, Breiman was already
interested in classification, co-authoring a paper on the convergence
properties of a ``Learning Algorithm'' \citep{LBandZW}. Curiously, the
paper had only two references, one of which was to some early work by
Seymour Papert, who was later to become one of the pioneers of
artificial intelligence and co-author of an influential (and
controversial) book on perceptrons \citep{minsky}.

After resigning, the first thing Breiman did was to write his
probability book \citep{probbook} and then, with no formal statistical
training, he proceeded to spend the next 13 years as a consultant.
As well as some work in transportation, he worked for William Meisel's
division of Technology Services Corporation, doing environmental
studies and unclassified defense work.
It's difficult to imagine making such a transition today, but one can
speculate that it was in part, \textit{because} he did not have a
background in applied statistics that Breiman was so successful at consulting.
Certainly the prediction problems on which he worked, some of which are
mentioned in \citet{Nailfinders} and Section 3 of \citet{twocultures},
would have been a challenge for the tools and computers of the time. In
\citet{twocultures}, he acknowledges Meisel for helping him ``make the
transition from probability theory to algorithms.''

\section{Classification and regression trees}

One of the early problems Breiman worked on as a consultant was to
classify ship types from the peaks of radar profiles. The observations
had different numbers of peaks and the number of peaks and their
locations depended on the angle the ship made with the radar. After ``a~lot of head-scratching and a lot of time just thinking'' the idea of a
classification tree came to him ``out of the blue.'' After this,
Meisel's research team began using trees regularly. Charles Stone was
brought on board, became interested in trees, and worked with Breiman
to improve accuracy. In the early to mid-1970s, Breiman and Stone came
up with the breakthrough idea of using cross validation to prune large trees.

It's difficult to obtain published work from Breiman's consulting
years, but
by 1976, Breiman and Meisel published an early version of regression
trees \citep{LBandWM} which broke down the data space into regions and
fitted a linear regression in each region. Regions were split using a
randomly oriented plane and an F-ratio was used to determine if the
split had significantly reduced the residual sum of squares; if not,
another random split was tried.
In retrospect, the idea of using randomly chosen splits seems a good 20
years ahead of its time. The statement
``many typical data analytic problems are characterized by their high
dimensionality\ldots and the lack of any  a priori  identification
of a natural and appropriate family of regression functions'' \citep
{LBandWM} was a clear indicator of Breiman's future research directions.

In 1976, Breiman met Jerome Friedman, a high-energy particle physicist,
and soon Friedman was also working as a consultant for TSI. Both
Friedman and Stone had connections to Richard Olshen, and the four
started to collaborate. Apparently, they decided to publish their
research as a book because they believed the work was unlikely to be
published in the standard statistical journals.

In 1980, Stone and Breiman joined the UC Berkeley Statistics
Department, and the group experimented with different splitting
criteria, refined the cross-validation approach, and came up with the
idea of surrogate splits. Several things set this work apart from other
early work on trees. First, they did painstaking experiments. As they
report in \citet{Lohcomment}, ``In the course of the research that led
to CART, almost two years were spent experimenting with different
stopping rules. Each stopping rule was tested on hundreds of simulated
data sets with different structures.'' Second, they kept applications
in the foreground of their work, due in part to Breiman's years as a
consultant. Third, they had what Breiman referred to as ``some
beautiful and complex theory.''
The book, priced low to make it accessible, was published in 1984 \citep{CART}.

\section{ACE and additive models}

I once heard Charles Stone express regret that the CART group had not
written a follow up book of ``things we tried that didn't work.'' I
expect such a book could have prevented a number of researchers from
reinventing the wheel, but few would want to read such a book, much
less write it.
In fact, after completing \citet{CART}, Breiman admitted to being
``completely fed up with thinking about trees.'' Breiman and Friedman
continued to talk, because both were interested in high-dimensional
data analysis, and soon they came up with the Alternating Conditional
Expectations (ACE) algorithm \citep{ACE}. For predictor variables
$X_1,\ldots,X_p$ and response $Y$,
ACE defines $\phi_1^{\star},\ldots ,\phi_p^{\star}$ and $\theta^{\star}$
to minimize
\[
\mathrm{E} \Biggl[ \theta(Y) - \sum_{j=1}^p \phi_j (X_j)  \Biggr]^2
\]
under the constraint $\operatorname{Var}(\theta) = 1$.
Estimates $\phi_1^{\star},\ldots ,\phi_p^{\star}$ and $\theta^{\star}$
were obtained using an iterative optimization procedure involving
(nonlinear) smoothing to estimate each of the transformations while
holding the others fixed. This was an application of the Gauss--Seidel
algorithm of numerical linear algebra. A simpler version, taking $\theta
$ as the identity, is the familiar ``backfitting'' algorithm [\citet{GAMS}, \citet{buja}].

ACE was the first in a series of papers Breiman wrote on smoothing and
additive models. \citet{LBPeters} compared four scatterplot smoothers
using an extensive simulation. Building on the spline models used in
\citet{LBPeters}, Breiman's $\Pi$ method \citep{Pimple}, with the
colorful acronym ``PIMPLE,'' fit additive models of products of
(univariate) cubic splines.
Hinging hyperplanes \citep{hinging} fit an additive function of
hyperplanes, continuously joined along a line called a ``hinge.''
According to \citet{additive}, while ACE provided the ``first available
method for fitting additive models to data,'' it had some difficulties.
For small sample sizes, the results were ``noisy and erratic.'' The
nonlinearity of the smoother combined with the iterative algorithm led
to results that were ``difficult to analyze and sometimes mildly
unstable.'' So Breiman went back to the drawing board, adapting a
spline-based method using stepwise deletion of knots \citep{psmith},
resulting in \citet{additive}. This paper contains early thoughts on
using cross-validation to measure instability: ``If transformations
change drastically when one or a few cases are removed, then they do
not reflect an overall pattern in the data.'' These early ideas of
instability ultimately led to some of Breiman's most influential work.

\section{Multivariate techniques}

While all Breiman's work was multivariate, some was more clearly
affiliated with traditional multivariate techniques. In 1984, Breiman
and Ihaka released a technical report \citep{LBandIHAKA} describing a
nonlinear, smoothing-based version of discriminant analysis. The work
was never published but it motivated the work on ``Flexible
Discriminant Analysis'' by \citet{FDA}.

In his consulting days, one of the problems Breiman studied was
next-day ozone prediction. One of his ideas was to represent each day
as a mixture of ``extreme'' or ``archetypal'' days. For example, an
archetypal sunny day would be as sunny as possible, an archetypal rainy
day would have as much rain as possible, an archetypal foggy day would
have fog for as long as possible, and so on. Most days would not be
archetypal---they would fall in between the archetypes, resembling each
to a greater or lesser extent. For data $\{ x_i, i=1,\ldots ,N\}$, the
problem was to find archetypal points $\{z_k, k=1,\ldots ,K \}$ to minimize
\[
\sum_i \biggl\| x_i - \sum_k \alpha_{ik} z_k \biggr\|^2
\]
subject to the constraints $\alpha_{ik} \geq0, \sum_k \alpha_{ik} =
1$, while also constraining the $z_k$'s to fall on or inside the convex
hull of the data. The problem can be solved using an alternating least
squares algorithm \citep{arche}. Archetypes have been used as an
alternative to cluster analysis or principal components in numerous disciplines.

The final method in this section is a paper on multivariate regression,
whimsically called ``curds and whey'' \citep{curds}. To predict
correlated responses, Breiman and Friedman considered predicting each
response by a linear combination of the ordinary least squares (OLS)
predictors rather than the OLS predictors themselves. The method worked
by transforming into canonical coordinates, shrinking, then
transforming back. Cross-validation was used to choose the amount of shrinkage.

\section{Subset selection in linear regression}

Breiman had a longstanding interest in submodel selection in linear
regression, revealing itself in \citet{LBandWM}, which used an early
version of a regression tree to estimate the ``intrinsic variability''
of the data, with the goal of effectively ranking the predictive
capabilities of subsets of independent variables. \citet{LBandDF}
looked at determining the optimal number of regressors to minimize mean
squared prediction error. Again, using prediction error as the gold
standard, \citet{littleb} and \citet{LBandPS} contained careful and
thorough simulation studies for the $X$-fixed and $X$-random situations.

As \citet{Efroncomment} mentioned, Leo's ``openness to new ideas
whatever their source'' was an attractive feature of his work. One
example of this openness was that
in the early 1990s, Leo got interested in neural nets and started
participating in the Neural Information Processing Systems (NIPS)
conference and workshops. Neural nets were not really a new idea,
but they were enjoying new popularity among computer scientists,
physicists and engineers, who in Leo's view were turning out
``thousands of interesting research papers related to applications and
methodology'' \citep{twocultures}. To this active community, Leo
brought his considerable statistical background, experience with trees
and subset selection, and perspective from years of dealing with real
data and thinking about how to do it better. This led to Leo's most
productive years, in part facilitated by his retirement from the UC
Berkeley Statistics Department in early 1993, about which he said, ``So
far retirement has meant that I've got more time to spend on research''
(personal communication).

The first work to appear from this period, stacking \citep{stacking},
was stimulated by \citet{Wolpert} and first appeared as a technical
report in 1992. In \citet{stacking}, he said, ``In past statistical
work, all the focus has been on selecting the ``best'' single model from
a class of models. We may need to shift our thinking to the possibility
of forming combinations of models.'' In the case of stacking, this was
a linear combination of predictors. Each predictor was based on what
Wolpert called the ``level 1 data'' \citep{Wolpert}. \citet{stacking}
considered a family of models indexed by $k=1,\ldots ,K$. For example,
$k$ might be the number of variables in a subset selection method or
$k$ might index a collection of shrinkage parameters $\{ \lambda_k,
k=1,\ldots ,K \}$ for ridge regression.
For data $\{x_{1n},\ldots ,x_{pn},y_n, n=1,\ldots ,N \}$, each of the $K$
predictors were fit to the data with observation $n$ omitted
(leave-one-out cross validation) to give $k$ predictions of $y_n$,
namely $z_{kn}, k=1,\ldots ,K$, which were the ``level 1 data.'' The
``stacked'' predictor was $\sum_k \alpha_k z_{kn}$ where $\alpha_k \geq
0, k=1,\ldots ,K$, were chosen to minimize
\[
\sum_n \biggl ( y_n - \sum_k \alpha_k z_{kn}  \biggr) ^2.
\]
Breiman considered stacked subsets and stacked ridge regressions and
concluded that both were better than the existing method (choosing a
single model by cross-validation). However, stacking improved subsets
more than it improved ridge, which Breiman suggested was due to the
greater instability of subset selection.

Building on stacking \citep{stacking} and using some of his experiences
from \citet{littleb} and \citet{LBandPS}, Breiman introduced the
nonnegative garrote \citep{garrote}. For data as before and original
OLS coefficients $\{\hat{\beta}_k\}$, the nonnegative garrote chose $\{
c_k \}$ to minimize
\[
\sum_n  \biggl( y_n - \sum_k c_k \hat{\beta}_k x_{kn}  \biggr)^2
\]
subject to the constraints $c_k \geq0$ and $\sum_k c_k \leq s$. This
was a much simpler idea than stacking because it did not use Wolpert's
``level 1 data'' \citep{Wolpert} and $k$ ranged over the predictor
variables instead of denoting the size of a subset or the value of a
shrinkage parameter. Breiman found \citep{garrote} that the garrote had
consistently lower prediction error than subset selection, and
sometimes better than ridge regression. Breiman's ideas about
instability, first mentioned in \citet{additive}, led him to
characterize of ridge regression as stable, subset selection unstable,
and the garrote intermediate. Breiman remarked that ``the more unstable
a procedure is, the more difficult it is to accurately estimate PE
(prediction error)'' and speculated about finding a ``numerical measure
of stability.'' \citet{sparseboost} showed some interesting results for
the garrote in a boosting context. However, the largest impact of the
garrote was that it inspired the lasso [\citet{lasso}], which is
currently the method of choice, in part because of garrote's dependence
on $\{\hat{\beta}_k\}$.

Breiman's notions of stability were further explored in \citet
{heuristics}. He compared ridge regression, subset selection and two
versions of garrote and stated, ``Unstable procedures can be stabilized
by perturbing the data, getting a new predictor sequence\ldots  and then
averaging over many such predictor sequences.'' The types of
perturbation he considerd are leave-one-out cross-validation,
leave-ten-out cross-validation and adding random noise to the response
variable. He stated \citep{heuristics} ``we do not know yet what the
best stabilization method is.''

\section{Bagging}

Breiman released an early version of \citet{heuristics} in June 1994,
but by September of the same year he released yet another technical
report in which he had already resolved some of the questions raised in
\citet{heuristics}. He called the report ``Bagging Predictors'' and it
was to be published as \citet{bagging}. The name comes from ``bootstrap
aggregating'' because in bagging, the data were perturbed by taking
bootstrap samples and the resulting predictors were averaged
(aggregated) to give the ``bagged estimate.'' The classification
version aggregates by voting the predictors. In November 1994, Breiman
presented bagging as part of a Tutorial at the NIPS conference, where
it was immediately embraced by the neural net community. According to
Google Scholar, citations of \citet{bagging} already exceed 6300,
slightly higher than Efron's 1979 bootstrap paper \citep{efron}. The
simplicity and elegance of bagging made it appealing in a community
where new ideas tended to be technically complex.

In bagging, each predictor was fit to a bootstrap sample, so roughly
37\% of the observations were not included in the fit (``out-of-bag'').
In an unpublished technical report \citet{oob} described how to use
these for estimating node probabilities and generalization error.

Although bagging trees improved the accuracy of trees, Breiman liked
the simple, understandable structure of individual trees and was not
ready to give up on them. Noting that trees have ``the disadvantage
that the splits get noisier as you go down'' (personal communication), he
worked with Nong Shang \citep{distributionbased} to try to improve the
stability of trees by estimating the joint density of the data and
basing the splits on this density estimate instead of directly on the
data. One of the problems of this method was that density estimates
depended on numerous parameters and Breiman referred to it later \citep
{convexpseudo} as a ``complex and unwieldy procedure.'' Another
attempt, described in \citet{convexpseudo}, was to generate new
``pseudo-data'' by randomly choosing an existing data point and moving
its predictor variables a small step towards a second randomly-chosen
data point. The new predictor values, together with the response for
the original data point, gave the pseudo-data. The step size was chosen
to be uniform on the interval $(0,d)$ where $d$ was a parameter of the
method. Although the results appeared promising, the method did not
give improvements on large datasets and the paper was never published.

Breiman tried to improve upon bagging in a number of other ways. His
``iterated'' or ``adaptive'' bagging \citep{itbag} was designed to
reduce the bias of bagged regressions by successively altering the
output values using the out-of-bag data. Naturally, this biases the
out-of-bag generalization error estimates, but Breiman found that for
the purpose of bias reduction it worked well \citep{itbag}. In a
similar vein, \citet{rout} provided an alternative to bagging by
combining predictors fit to data for which only the output variables
have been perturbed. It's not clear whether these ideas would have
endured because Breiman did not release code and they were discarded
once he discovered random forests \citep{RF}.

\section{Boosting and arcing}

While Breiman developed bagging, Freund and Schapire worked on AdaBoost
[\citet{ada1}, \citet{ada2}, \citet{ada3}].
Breiman referred to the AdaBoost algorithm as ``the most accurate
general purpose classification algorithm available'' \citep{WaldAnnals}.
Like bagging, AdaBoost combined a sequence of predictors. Unlike
bagging, each predictor was fit to a sample from the training data,
with larger sampling weights given to observations that had been
misclassified by earlier predictors in the sequence. The predictions
were combined using performance weights.
In a personal communication, Breiman wrote, ``Some of my latest efforts
are to understand Adaboost
better. Its really a strange algorithm with unexpected behavior.
Its become like searching for the Holy Grail!!''
In his quest, Breiman produced a series of papers [\citeauthor{arcingclassifiers}
(\citeyear{arcingedge,arcingclassifiers,halfhalf,predictiongames,infinitytheory,WaldAnnals})]. He noted in \citet{arcingclassifiers} that if AdaBoost
``is run far past the point at which the training set error is zero, it
gives better performance than bagging on a number of real data sets.''
This was a great mystery and Breiman was determined to get to the
bottom of it. In \citet{arcingclassifiers}, Breiman constructed a more
general class of algorithms ``arcing,'' of which AdaBoost, (``arc-fs'')
was a special case. One contribution of \citet{arcingclassifiers} was
that Breiman removed the randomness of boosting by using a weighted
version of the classifier instead of sampling weights.
Focusing on bias and variance, he concluded that ``Arcing does better
than bagging because it does better at variance reduction'' \citep
{arcingclassifiers}, but \citet{margin} gave examples in which the
main effect of AdaBoost was to reduce bias and proposed their own
reasons for why boosting worked so well. Breiman thought the
explanation was incomplete \citep{predictiongames}.

Breiman's work on half and half bagging \citep{halfhalf} was stimulated
by one of the referees of \citet{arcingclassifiers}, who commented that
the probability weight at a given step was equally divided between the
points misclassified, and those correctly classified, at the previous
step. In \citet{halfhalf} Breiman divided the data into two parts, one
containing ``easy'' points, the other ``hard'' points, based on
previous classifiers in the sequence. He randomly sampled an equal
number of cases from both groups and fitted a classification tree. For
the first time, the tree was grown deep (one example per terminal
node), which he later carried over to random forests \citep{RF}.

In \citet{arcingedge}, he showed that AdaBoost is a
``down-the-gradient'' method for minimizing an exponential function of
the error. Independently,
\citet{additivelogistic} presented ``The Statistical View of Boosting.''

About his ``Infinity Theory'' paper \citep{infinitytheory}, Breiman
stated in August 2000:
``I've been compulsively working on a theory paper about tree ensembles
which I got sick and tired
of but knew that if I didn't keep going\ldots  it would never get finished.''
The paper was released as a technical report, cited by \citet
{tongzhang} and \citet{boostl2}, among others. A later version was
published as \citet{WaldAnnals} and in this paper Breiman showed that
the population version of AdaBoost was Bayes-consistent. In the
meantime, several publications, including \citet{additivelogistic},
suggested that AdaBoost could overfit in the limit and \citet{jiang}
showed that in the finite sample case, AdaBoost was only
Bayes-consistent if it was regularized.

\section{Random forests}

In the light of boosting, Breiman spent a lot of time trying to improve
individual trees [\citet{distributionbased}, \citet{convexpseudo}] and
bagged trees [\citeauthor{rout} (\citeyear{rout,itbag})]. He also worked very hard to
understand what was going on with boosting
[\citeauthor{arcingedge} (\citeyear{arcingedge,arcingclassifiers,halfhalf,predictiongames,infinitytheory,WaldAnnals})]. However, he never seriously
produced a boosting algorithm for practical use, and I believe the
reason was that he wanted a method that could give meaningful results
for data analysis, not just prediction, and he didn't think he could
get this by combining dependent predictors.
The culmination of his work on bagging and how to improve it, and his
work trying to understand boosting, was a method Breiman called
``random forests'' (RF) \citep{RF}. Random forests fit trees to
independent bootstrap samples from the data. The trees were grown large
(for classification) and at each node independently, $m$ predictors
were chosen out of the $p$ available, and the best possible split on
these $m$ predictors was used. As a default for classification, Breiman
settled on choosing $m=\sqrt{p}$. In RF we see a synthesis of the
bagging ideas (bootstrapping), along with ideas that came from boosting
(growing large trees), and Breiman's understanding of how to increase
instability (randomly choosing predictors at each node) to get more
accurate aggregate predictions. Once he came up with RF, Breiman
stopped working on new algorithms and started work on how to get the
most out of the RF results. He developed measures of variable
importance and proximities between observations. Together, we developed
a program for visualizing and interpreting RF results (available from
\url{http://www.math.usu.edu/~adele/forests/cc\_graphics.htm}). Chao
Chen and Andy Liaw worked with Breiman on ways to adjust RF for
unbalanced classes \citep{imbalanced}. Vivian Ng worked with him on
detecting interactions \citep{Vivian}. In his last technical report,
Breiman showed consistency for a simple version of RF [\citet{consistency}].
But the work on RF did not stop when Breiman died. Several extensions
have been published; for example, \citet{diaz} developed a variable
selection procedure, \citet{Meinshausen} introduced quantile
regression forests, and \citet{sandrine}, \citet{ishwaran} considered
forests for survival analysis. Although theory is still thin on the
ground, \citet{linjeon} showed that RF behaves like a nearest neighbor
classifier with an adaptive metric and Biau, Devroye and Lugosi made some progress
on consistency in a paper dedicated to Breiman's memory \citep{Biau}.
Numerous applied articles have appeared and even a number of YouTube
videos. I believe Breiman would be truly delighted at the popularity of
the method.

\section{Software}

Leo developed his own code, invariably in fortran. I collaborated with
him on the random forests fortran code and documentation
\url{http://www.math.usu.edu/~adele/forests/cc\_home.htm}.
Andy Liaw and Matt Wiener developed an interface to R \citep{Andy}.
Although Leo supported the R release and admired the free-software
philosophy of R, he regarded R as a tool for ``Ph.D. statisticians'' and
he wanted his code to also be available with an easy to use graphical
user interface (GUI). GUI-driven software for classification and
regression trees and random forests is available from Salford Systems.
Versions of trees, random forests and archetypes are available in R
(packages rpart, randomForests \citep{Andy}, and archetypes \citep{archetypes}).

\section{Textbooks}

In addition to his papers, Breiman wrote three textbooks [\citeauthor{probbook}
(\citeyear{probbook,eprobbook,statbook})], the first of which is in SIAM's
``Classics of Mathematics'' series. Perhaps even more impressive is the
fact that other scholars are now writing texts that refer extensively
to Breiman's work, including trees, bagging and random forests [see
\citet{Berk}, \citet{HTF} and \citet{Izenman}].

\section{Philosophy}

Breiman passionately believed that statistics should be motivated by
problems in data analysis. Comments such as

\begin{quote}

If statistics is an applied field and not a minor branch of
mathematics, then more than 99\% of the published papers are useless
exercises. \citep{Wolpertbook}

\end{quote}

\noindent
show how deeply he believed that statistics needed a change of
direction. When he heard that \citet{arcingclassifiers} was to be
published with discussion in \textit{The Annals of Statistics}, he
commented that ``it would sure liven things up\ldots  maybe get some blood
moving in the statistical main stream of asymptopia'' (personal communication).

Although it is not widely cited, I believe Breiman's ``Two Cultures''
paper \citep{twocultures} is one of his most widely read, at least
among statisticians. The paper contained Breiman's views about where
the field was going and what needed to be done. To conclude, he said:

\begin{quote}

The roots of statistics, as in science, lie in working
with data and checking theory against data. I
hope in this century our field will return to its roots.
There are signs that this hope is not illusory. Over
the last ten years, there has been a noticeable move
toward statistical work on real world problems and
reaching out by statisticians toward collaborative
work with other disciplines. I believe this trend will
continue and, in fact, has to continue if we are to
survive as an energetic and creative field. [\citet{twocultures}]

\end{quote}


\printaddresses


\begin{thebibliography}{99}
\bibitem[\protect\citeauthoryear{Berk}{2008}]{Berk}
\textsc{Berk, R.} (2008).
\textit{Statistical Learning from a Regression Perspective.}
Springer, New York.

\bibitem[\protect\citeauthoryear{Biau, Devroye and Lugosi}{2008}]{Biau}
\textsc{Biau, G., Devroye, L.}
and
\textsc{Lugosi, G.} (2008).
Consistency of random forests and other averaging classifiers.
\textit{J. Mach. Learn. Res.}
\textbf{9} 2039--2057.
\MR{2447310}

\bibitem[\protect\citeauthoryear{Breiman}{1968}]{probbook}
\textsc{Breiman, L.} (1968).
\textit{Probability Theory.} Addison-Wesley, Reading, MA.
[Republished (1991) in \textit{Classics of Mathematics.} SIAM,
Philadelphia, PA.]
\MR{0229267}

\bibitem[\protect\citeauthoryear{Breiman}{1969}]{eprobbook}
\textsc{Breiman, L.} (1969).
\textit{Probability and Stochastic Processes with a View Toward Applications.}
Houghton Mifflin, Boston, MA.
\MR{0254942}

\bibitem[\protect\citeauthoryear{Breiman}{1973}]{statbook}
\textsc{Breiman, L.} (1973).
\textit{Statistics: With a View Toward Applications.} Houghton Mifflin
Harcourt, Boston, MA.
\MR{0359089}

\bibitem[\protect\citeauthoryear{Breiman}{1984}]{Nailfinders}
\textsc{Breiman, L.} (1984).
Nail finders, edifices, and Oz.
Technical Report 32. Dept. Statistics, Univ. California,
Berkeley, CA. Neyman--Kiefer Memorial Volume.

\bibitem[\protect\citeauthoryear{Breiman}{1991}]{Pimple}
\textsc{Breiman, L.} (1991).
The $\Pi$ method for estimating multivariate functions from noisy data.
\textit{Technometrics}
\textbf{33} 125--143.
\MR{1110355}

\bibitem[\protect\citeauthoryear{Breiman}{1992}]{littleb}
\textsc{Breiman, L.} (1992).
The little bootstrap and other methods for
dimensionality selection in regression: $X$-fixed prediction error.
\textit{J. Amer. Statist. Assoc.}
\textbf{87} 738--754.
\MR{1185196}

\bibitem[\protect\citeauthoryear{Breiman}{1993a}]{additive}
\textsc{Breiman, L.} (1993a).
Fitting additive models to regression data: Diagnostics and alternative views.
\textit{J. Comput. Statist. Data Anal.}
\textbf{15} 13--46.
\MR{1202297}

\bibitem[\protect\citeauthoryear{Breiman}{1993b}]{hinging}
\textsc{Breiman, L.} (1993b).
Hinging hyperplanes for regression, classification, and function approximation.
\textit{IEEE Trans. Inform. Theory}
\textbf{39} 999--1013.
\MR{1237723}

\bibitem[\protect\citeauthoryear{Breiman}{1994}]{census}
\textsc{Breiman, L.} (1994).
The 1991 census adjustment: Undercount or bad data?
\textit{Statist. Sci.}
\textbf{9} 458--537.

\bibitem[\protect\citeauthoryear{Breiman}{1995a}]{garrote}
\textsc{Breiman, L.} (1995a).
Better subset regression using the nonnegative garrote.
\textit{Technometrics}
\textbf{37} 373--384.
\MR{1365720}

\bibitem[\protect\citeauthoryear{Breiman}{1995b}]{Wolpertbook}
\textsc{Breiman, L.} (1995b).
Reflections after refereeing papers for NIPS.
In \textit{The Mathematics of Generalization: Proceedings of the SFI/CNLS
Workshop on Formal Approaches to Supervised Learning, Volume 1992}
(D. H. Wolpert, ed.).
Westview Press, Boulder, CO.
\MR{1353248}

\bibitem[\protect\citeauthoryear{Breiman}{1996a}]{stacking}
\textsc{Breiman, L.} (1996a).
Stacked regressions.
\textit{Mach. Learn.}
\textbf{24} 49--64.

\bibitem[\protect\citeauthoryear{Breiman}{1996b}]{heuristics}
\textsc{Breiman, L.} (1996b).
Heuristics of instability and stabilization in model selection.
\textit{Ann. Statist.}
\textbf{24} 2350--2383.
\MR{1425957}

\bibitem[\protect\citeauthoryear{Breiman}{1996c}]{bagging}
\textsc{Breiman, L.} (1996c).
Bagging predictors.
\textit{Mach. Learn.}
\textbf{24} 123--140.

\bibitem[\protect\citeauthoryear{Breiman}{1997a}]{arcingedge}
\textsc{Breiman, L.} (1997a).
Arcing the edge.
Technical Report 486. Dept. Statistics, Univ. California, Berkeley, CA.

\bibitem[\protect\citeauthoryear{Breiman}{1997b}]{oob}
\textsc{Breiman, L.} (1997b).
Out-of-bag estimation. Technical report. Dept. Statistics,
Univ. California, Berkeley, CA.

\bibitem[\protect\citeauthoryear{Breiman}{1998a}]{arcingclassifiers}
\textsc{Breiman, L.} (1998a).
Arcing classifiers.
\textit{Ann. Statist.}
\textbf{26} 801--849.
\MR{1635406}

\bibitem[\protect\citeauthoryear{Breiman}{1998b}]{convexpseudo}
\textsc{Breiman, L.} (1998b).
Using convex pseudo-data to increase prediction
accuracy. Technical Report~513. Dept. Statistics, Univ. California,
Berkeley, CA.

\bibitem[\protect\citeauthoryear{Breiman}{1998c}]{halfhalf}
\textsc{Breiman, L.} (1998c).
Half \& half bagging and hard boundary points. Technical Report~534. Dept.
Statistics, Univ. California, Berkeley, CA.

\bibitem[\protect\citeauthoryear{Breiman}{1999a}]{predictiongames}
\textsc{Breiman, L.} (1999a).
Prediction games and arcing algorithms.
\textit{Neural Comput.}
\textbf{11} 1493--1517.

\bibitem[\protect\citeauthoryear{Breiman}{1999b}]{pasting}
\textsc{Breiman, L.} (1999b).
Pasting small votes for classification in large databases and on-line.
\textit{Mach. Learn.}
\textbf{36} 85--103.

\bibitem[\protect\citeauthoryear{Breiman}{2000a}]{rout}
\textsc{Breiman, L.} (2000a).
Randomizing outputs to increase prediction accuracy.
\textit{Mach. Learning}
\textbf{40} 229--242.

\bibitem[\protect\citeauthoryear{Breiman}{2000b}]{infinitytheory}
\textsc{Breiman, L.} (2000b).
Some infinity theory for predictor ensembles. Technical Report~577.
Dept. Statistics, Univ. California, Berkeley, CA.

\bibitem[\protect\citeauthoryear{Breiman}{2001a}]{RF}
\textsc{Breiman, L.} (2001a).
Random forests.
\textit{Mach. Learn.}
\textbf{45} 5--32.

\bibitem[\protect\citeauthoryear{Breiman}{2001b}]{itbag}
\textsc{Breiman, L.} (2001b).
Using iterated bagging to debias regressions.
\textit{Mach. Learn.}
\textbf{45} 261--277.

\bibitem[\protect\citeauthoryear{Breiman}{2001c}]{twocultures}
\textsc{Breiman, L.} (2001c).
Statistical modeling: The two cultures.
\textit{Statist. Sci.}
\textbf{16} 199--231.
\MR{1874152}

\bibitem[\protect\citeauthoryear{Breiman}{2004a}]{consistency}
\textsc{Breiman, L.} (2004a).
Consistency for a simple model of random forests.
Technical Report~670. Dept. Statistics, Univ. California, Berkeley, CA.

\bibitem[\protect\citeauthoryear{Breiman}{2004b}]{WaldAnnals}
\textsc{Breiman, L.} (2004b).
Population theory for boosting ensembles.
\textit{Ann. Statist.}
\textbf{32} 1--11.
\MR{2050998}

\bibitem[\protect\citeauthoryear{Breiman and Cutler}{1993}]{global}
\textsc{Breiman, L.}
and
\textsc{Cutler, A.} (1993).
A deterministic algorithm for global optimization.
\textit{Math. Programming}
\textbf{58} 179--199.
\MR{1216490}

\bibitem[\protect\citeauthoryear{Breiman and Freedman}{1983}]{LBandDF}
\textsc{Breiman, L.}
and
\textsc{Freedman, D.} (1983).
How many variables should be entered in a regression equation?
\textit{J. Amer. Statist. Assoc.}
\textbf{78} 131--136.
\MR{0696857}

\bibitem[\protect\citeauthoryear{Breiman and Friedman}{1985}]{ACE}
\textsc{Breiman, L.}
and
\textsc{Friedman, J.} (1985).
Estimating optimal transformations for multiple regression and correlation.
\textit{J. Amer. Statist. Assoc.}
\textbf{80} 580--598.
\MR{0803258}

\bibitem[\protect\citeauthoryear{Breiman and Friedman}{1988}]{Lohcomment}
\textsc{Breiman, L.}
and
\textsc{Friedman, J.} (1988).
Comment on ``Tree-structured classification via generalized
discriminant analysis'' by W. Y. Loh and N. Vanichsetakul.
\textit{J. Amer. Statist. Assoc.}
\textbf{83} 725--727.
\MR{0963799}

\bibitem[\protect\citeauthoryear{Breiman and Friedman}{1997}]{curds}
\textsc{Breiman, L.}
and
\textsc{Friedman, J.} (1997).
Predicting multivariate responses in multiple linear regression.
\textit{J. Roy. Statist. Soc. Ser. B}
\textbf{59} 3--54.
\MR{1436554}

\bibitem[\protect\citeauthoryear{Breiman et al.}{1984}]{CART}
\textsc{Breiman, L., Friedman, J., Olshen, R.}
and
\textsc{Stone, C.} (1984).
\textit{Classification and Regression Trees.}
Wadsworth, New York.
\MR{0726392}

\bibitem[\protect\citeauthoryear{Breiman and Ihaka}{1984}]{LBandIHAKA}
\textsc{Breiman, L.}
and
\textsc{Ihaka, R.} (1984).
Nonlinear discriminant analysis via scaling and ACE. Technical
Report~40. Dept. Statistics, Univ. California,
Berkeley, CA.

\bibitem[\protect\citeauthoryear{Breiman and Meisel}{1976}]{LBandWM}
\textsc{Breiman, L.}
and
\textsc{Meisel, W. S.} (1976).
General estimates of the intrinsic variability of data in nonlinear
regression models.
\textit{J. Amer. Statist. Assoc.}
\textbf{71} 301--307.

\bibitem[\protect\citeauthoryear{Breiman, Meisel and Purcell}{1977}]{Varkern}
\textsc{Breiman, L., Meisel, W. S.}
and
\textsc{Purcell, E.} (1977).
Variable kernel estimates of multivariate densities.
\textit{Technometrics}
\textbf{19} 135--144.

\bibitem[\protect\citeauthoryear{Breiman and Peters}{1992}]{LBPeters}
\textsc{Breiman, L.}
and
\textsc{Peters, S.} (1992).
Comparing automatic smoothers (A public service enterprise).
\textit{Int. Statist. Rev.}
\textbf{60} 271--290.

\bibitem[\protect\citeauthoryear{Breiman and Spector}{1992}]{LBandPS}
\textsc{Breiman, L.}
and
\textsc{Spector, P.} (1992).
Submodel selection and evaluation in regression. The \textit{X}-random case.
\textit{Int. Statist. Rev.}
\textbf{60} 291--319.

\bibitem[\protect\citeauthoryear{Breiman, Tsur and Zemel}{1993}]{censreg}
\textsc{Breiman, L., Tsur, Y.}
and
\textsc{Zemel, A.} (1993).
On a simple estimation procedure for censored regression models with
known error distributions.
\textit{Ann. Statist.}
\textbf{21} 1711--1720.
\MR{1245765}

\bibitem[\protect\citeauthoryear{Breiman and Wurtele}{1964}]{LBandZW}
\textsc{Breiman, L.}
and
\textsc{Wurtele, Z. S.} (1964).
Convergence properties of a learning algorithm.
\textit{Ann. Math. Statist.}
\textbf{35} 1819--1822.
\MR{0178995}

\bibitem[\protect\citeauthoryear{B\"{u}hlmann and Yu}{2003}]{boostl2}
\textsc{B\"{u}hlmann, P.}
and
\textsc{Yu, B.} (2003).
Boosting with the L2 loss: Regression and classification.
\textit{J.~Amer. Statist. Assoc.}
\textbf{98} 324--339.
\MR{1995709}

\bibitem[\protect\citeauthoryear{B\"{u}hlmann and Yu}{2006}]{sparseboost}
\textsc{B\"{u}hlmann, P.}
and
\textsc{Yu, B.} (2006).
Sparse boosting.
\textit{J. Mach. Learn. Res.}
\textbf{7} 1001--1024.
\MR{2274395}

\bibitem[\protect\citeauthoryear{Buja, Hastie and Tibshirani}{1989}]{buja}
\textsc{Buja, A., Hastie, T.}
and
\textsc{Tibshirani, R.} (1989).
Linear smoothers and additive models.
\textit{Ann. Statist.}
\textbf{17} 453--510.
\MR{0994249}

\bibitem[\protect\citeauthoryear{Chao, Liaw and Breiman}{2004}]{imbalanced}
\textsc{Chao, C., Liaw, A.}
and
\textsc{Breiman, L.} (2004).
Using random forests to learn imbalanced data. Technical Report~666.
Dept. Statistics, Univ. California, Berkeley, CA.

\bibitem[\protect\citeauthoryear{Cutler and Breiman}{1994}]{arche}
\textsc{Cutler, A.}
and
\textsc{Breiman, L.} (1994).
Archetypal analysis.
\textit{Technometrics}
\textbf{36} 338--347.
\MR{1304898}

\bibitem[\protect\citeauthoryear{Diaz-Uriarte and Alvarez de
Andres}{2006}]{diaz}
\textsc{Diaz-Uriarte, R.}
and
\textsc{Alvarez de Andres, S.} (2006).
Gene selection and classification of microarray data using random forest.
\textit{BMC Bioinformatics}
\textbf{7} 3.

\bibitem[\protect\citeauthoryear{Efron}{1979}]{efron}
\textsc{Efron, B.} (1979).
Bootstrap methods: Another look at the jackknife.
\textit{Ann. Statist.}
\textbf{7} 1--26.
\MR{0515681}

\bibitem[\protect\citeauthoryear{Efron}{2001}]{Efroncomment}
\textsc{Efron, B.} (2001).
Comment on ``Statistical modeling: The two cultures'' by L. Breiman.
\textit{Statist. Sci.}
\textbf{16} 218--219.
\MR{1861072}

\bibitem[\protect\citeauthoryear{Eugster and Leisch}{2009}]{archetypes}
\textsc{Eugster, M. J. A.}
and
\textsc{Leisch, F.} (2009).
From Spider-Man to Hero---archetypal analysis in R.
\textit{J. Statist. Soft.}
\textbf{30} 1--23.

\bibitem[\protect\citeauthoryear{Freund}{1995}]{ada2}
\textsc{Freund, A.} (1995).
Boosting a weak learning algorithm by majority.
\textit{Inform. Comput.}
\textbf{121} 256--285.
\MR{1348530}

\bibitem[\protect\citeauthoryear{Freund and Schapire}{1996}]{ada3}
\textsc{Freund, Y.}
and
\textsc{Schapire, R.} (1996).
Experiments with a new boosting algorithm. In
\textit{Machine Learning: Proceedings of the Thirteenth International
Conference}
148--156. Morgan Kauffman, San Francisco, CA.

\bibitem[\protect\citeauthoryear{Hastie and Tibshirani}{1986}]{GAMS}
\textsc{Hastie, T.}
and
\textsc{Tibshirani, R.} (1986).
Generalized additive models.
\textit{Statist. Sci.}
\textbf{1} 297--310.
\MR{0858512}

\bibitem[\protect\citeauthoryear{Hastie, Tibshirani and Buja}{1994}]{FDA}
\textsc{Hastie, T., Tibshirani, R.}
and
\textsc{Buja, A.} (1994).
Flexible discriminant analysis by optimal scoring.
\textit{J. Amer. Statist. Assoc.}
\textbf{89} 1255--1270.
\MR{1310220}

\bibitem[\protect\citeauthoryear{Hastie, Tibshirani and
Friedman}{2000}]{additivelogistic}
\textsc{Hastie, T., Tibshirani, R.}
and
\textsc{Friedman, J.} (2000).
Additive logistic regression: A statistical
view of boosting (with discussion and a rejoinder by the authors).
\textit{Ann. Statist.}
\textbf{28} 337--407.
\MR{1790002}

\bibitem[\protect\citeauthoryear{Hastie, Tibshirani and Friedman}{2009}]{HTF}
\textsc{Hastie, T., Tibshirani, R.}
and
\textsc{Friedman, J.} (2009).
\textit{The Elements of Statistical Learning: Data Mining, Inference,
and Prediction},
2nd ed. Springer, New York.
\MR{1851606}

\bibitem[\protect\citeauthoryear{Hothorn et al.}{2006}]{sandrine}
\textsc{Hothorn, T., B\"{u}hlmann, P., Dudoit, S., Molinaro, A. }
and
\textsc{Van Der Laan, M.} (2006).
Survival ensembles.
\textit{Biostatistics}
\textbf{7} 355--373.

\bibitem[\protect\citeauthoryear{Ishwaran et al.}{2008}]{ishwaran}
\textsc{Ishwaran, H., Kogalur, U.~B., Blackstone, E.~H.}
and
\textsc{Lauer, M.~S.} (2008).
Random survival forests.
\textit{Ann. Appl. Statist.}
\textbf{2} 841--860.
\MR{2516796}

\bibitem[\protect\citeauthoryear{Izenman}{2008}]{Izenman}
\textsc{Izenman, A.} (2008).
\textit{Modern Multivariate Statistical Techniques.}
Springer, New York.
\MR{2445017}

\bibitem[\protect\citeauthoryear{Jiang}{2004}]{jiang}
\textsc{Jiang, W.} (2004).
Process consistency for AdaBoost.
\textit{Ann. Statist.}
\textbf{32} 13--29.
\MR{2050999}

\bibitem[\protect\citeauthoryear{Liaw and Wiener}{2002}]{Andy}
\textsc{Liaw, A.}
and
\textsc{Wiener, M.} (2002).
Classification and regression by random forest.
\textit{R News}
\textbf{2} 18--22.

\bibitem[\protect\citeauthoryear{Lin and Jeon}{2006}]{linjeon}
\textsc{Lin, Y.}
and
\textsc{Jeon, Y.} (2006).
Random forests and adaptive nearest neighbors.
\textit{J. Amer. Statist. Assoc.}
\textbf{101} 578--590.
\MR{2256176}

\bibitem[\protect\citeauthoryear{Meinshausen}{2006}]{Meinshausen}
\textsc{Meinshausen, N.} (2006).
Quantile regression forests.
\textit{J. Mach. Learn. Res.}
\textbf{7} 983--999.
\MR{2274394}

\bibitem[\protect\citeauthoryear{Minsky and Papert}{1969}]{minsky}
\textsc{Minsky, M.}
and
\textsc{Papert, P.} (1969).
\textit{Perceptrons: An Introduction to Computational Geometry.}
MIT Press, Cambridge, MA.

\bibitem[\protect\citeauthoryear{Ng and Breiman}{2005}]{Vivian}
\textsc{Ng, V.~W.}
and
\textsc{Breiman, L.} (2005).
Bivariate variable selection for classification problem.
Technical Report~692. Dept. Statistics, Univ. California, Berkeley, CA.

\bibitem[\protect\citeauthoryear{Olshen}{2001}]{Olshenbio}
\textsc{Olshen, R.} (2001).
A conversaton with Leo Breiman.
\textit{Statist. Sci.}
\textbf{16} 184--198.
\MR{1861072}

\bibitem[\protect\citeauthoryear{Schapire}{1990}]{ada1}
\textsc{Schapire, R.} (1990).
The strength of weak learnability.
\textit{Mach. Learn.}
\textbf{5} 197--227.

\bibitem[\protect\citeauthoryear{Schapire et al.}{1998}]{margin}
\textsc{Schapire, R., Freund, Y., Bartlett, P.}
and
\textsc{Lee, W.} (1998).
Boosting the margin: A new explanation for the effectiveness of voting methods.
\textit{Ann. Statist.}
\textbf{26} 1651--1686.
\MR{1673273}

\bibitem[\protect\citeauthoryear{Shang and Breiman}{1996}]{distributionbased}
\textsc{Shang, N.}
and
\textsc{Breiman, L.} (1996).
Distribution based trees are more accurate.
In \textit{Proceedings of the Int. Conf. on Neural Information
Processing, Hong Kong}
133--138.
Springer, Singapore.

\bibitem[\protect\citeauthoryear{Smith}{1982}]{psmith}
\textsc{Smith, P.} (1982).
Curve fitting and modeling with splines using statistical
variable selection techniques. NASA Report 166034. NASA,
Langley Research Center, Hampton, VA.

\bibitem[\protect\citeauthoryear{Tibshirani}{1996}]{lasso}
\textsc{Tibshirani, R.} (1996).
Regression shrinkage and selection via the lasso.
\textit{J. Roy. Statist. Soc. Ser. B}
\textbf{58} 267--288.
\MR{1379242}

\bibitem[\protect\citeauthoryear{Wolpert}{1992}]{Wolpert}
\textsc{Wolpert, D.} (1992).
Stacked generalization.
\textit{Neural Networks}
\textbf{5} 241--259.

\bibitem[\protect\citeauthoryear{Zhang}{2004}]{tongzhang}
\textsc{Zhang, T.} (2004).
Statistical behavior and consistency
of classification methods based on convex risk minimization.
\textit{Ann. Statist.}
\textbf{32} 56--134.
\MR{2051001}

\end{thebibliography}
\end{document}